\documentclass[]{aa}  
\usepackage{graphicx}
\usepackage{txfonts}
\usepackage{bm}

\newcommand{\unit}[1]{\ensuremath{\,\mathrm{#1}}}
\begin{document} 

   \title{Transition region adaptive conduction (TRAC) in multidimensional magnetohydrodynamic simulations}
   \subtitle{}
   \author{Yu-Hao Zhou
          \inst{1}
          \and
          Wen-Zhi Ruan
          \inst{1}
          \and
          Chun Xia
          \inst{2}
          \and
          Rony Keppens \inst{1}
          }
   \institute{Centre for mathematical Plasma Astrophysics, Department of Mathematics, KU Leuven, \\ Celestijnenlaan 200B, 3001 Leuven, Belgium.
              \email{yuhao.zhou@kuleuven.be}
         \and
             School of Physics and Astronomy, Yunnan University, Kunming 650050, China\\
             }
   \date{}
   \abstract
   {In solar physics, a severe numerical challenge for modern simulations is properly representing a transition region between the million-degree hot corona and a much cooler plasma of about 10000 K (e.g., the upper chromosphere or a prominence).  In previous 1D hydrodynamic simulations, the transition region adaptive conduction (TRAC) method has been proven to capture aspects better that are related to mass evaporation and energy exchange.}
   {We aim to extend this method to fully multidimensional magnetohydrodynamic (MHD) settings, as required for any realistic application in the solar atmosphere. Because modern MHD simulation tools efficiently exploit parallel supercomputers and can handle automated grid refinement, we design strategies for any-dimensional block grid-adaptive MHD simulations.}
   {We propose two different strategies and demonstrate their working with our open-source {\tt MPI-AMRVAC} code. We benchmark both strategies on 2D prominence formation based on the evaporation--condensation scenario, where chromospheric plasma is evaporated through the transition region and then is collected and ultimately condenses in the corona.}
   {A field-line-based TRACL method and a block-based TRACB method are introduced and compared in block grid-adaptive 2D MHD simulations. Both methods yield similar results and are shown to satisfactorily correct the underestimated chromospheric evaporation, which comes from a poor spatial resolution in the transition region.}
   {Because fully resolving the transition region in multidimensional MHD settings is virtually impossible, TRACB or TRACL methods will be needed in any 2D or 3D simulations involving transition region physics. }
   \keywords{Sun: Corona --
                Sun: magnetic fields --
                magnetohydrodynamics (MHD) --
                Sun: transition region -- 
                Methods: numerical
               }
   \maketitle
%
\section{Introduction}
\label{sec1}
Numerical tools for solving magnetohydrodynamic (MHD) equations play an important role in solar and astrophysics today. In the past decades, MHD simulations achieved remarkable progress through developments in high-performance computing and in the design of efficient and accurate shock-capturing solvers. All types of phenomena such as sunspots \citep{remp2009, remp2011}, solar prominences \citep{xia2012, xia2016}, coronal rain \citep{fang2013, fang2015, xia2017}, and solar flares \citep{cheu2019, ruan2020} can be simulated successfully. 
In these examples, multidimensional MHD simulations account for anisotropic thermal conduction and radiative cooling in the upper tenuous solar atmosphere, and incorporating these effects requires specific algorithmic approaches. The numerical strategy chosen to handle these effects also translates into constraints that limit affordable resolutions. 
At the same time, incorporating these effects is expected to be more realistic or reliable than assuming simpler zero-beta, isothermal, or purely ideal MHD conditions.

To handle more realistic dynamics in the solar atmosphere from the upper chromosphere to the corona, the governing equations of ideal MHD must at least be extended by additional source and sink terms for field-aligned thermal conduction and for (optically thin) radiative losses. In this approach, they appear only in the (total or internal) energy balance. It is difficult to include these terms in simulations because these source terms, especially the thermal conduction term, alter the purely hyperbolic character of the overall set of partial differential equations. The ideal MHD set only needs to account for the fastest wave speed, that is, the fast magnetosonic wave, in setting a time-step limit directly based on the chosen grid spacing. The thermal conduction introduces a parabolic diffusion-like source term, while optically thin radiative losses in essence introduce the light speed (or an infinite speed) in the system.
They have to be solved in an alternative way in order to prevent a too stringent limit on the allowed time step.
Many efforts have been introduced to accelerate and stabilize their calculation while ensuring a correct result.
Because it is an explicit approach, the super-time-stepping (STS) scheme is favored in handling thermal conduction, especially for large-scale simulations of clusters.
This STS method allows the thermal conduction term to advance forward by several super time-steps in a single MHD time step.
The restrictive Courant-Friedrichs-Lewy (CFL) condition can be relaxed in these super time-steps.
There are two variants of the STS method, the Runge-Kutta-Chebyshev STS (RKC-STS) method \citep{vand1980, verw1990, alex1996}, and the recently proposed Runge-Kutta-Legendre STS (RKL-STS) method \citep{meye2012, meye2014}.
The latter is considered to be more attractive and stable for simulating anisotropic terms on clusters \citep{vaid2017}.
Currently, STS methods have been implemented into many MHD codes, such as Athena++ \citep{ston2020}, Bifrost \citep{nobr2020}, and our open-source\footnote{See \url{http://amrvac.org}.} {\tt MPI-AMRVAC} code \citep{xia2018}.  For the optically thin radiative loss treatment, the exact integration scheme \citep{town2009} proved to be both efficient and robust compared with traditional fully explicit or (semi-)implicit schemes \citep{vanm2011}.

However, even with the modern advanced algorithmic methods for dealing with radiative losses and conduction, we still have to be very careful when these source terms need to be treated.
This is because they are still difficult to include, especially when a steep temperature gradient is involved.
In solar physics, the transition region (TR), a thin layer between the cool chromosphere and the hot corona, might be the most concerned. Similarly, whenever a coronal condensation forms by means of thermal instability \citep[e.g., ][]{fang2015, xia2016, clae2020}, the dynamically created boundary between the prominence or coronal rain blob and the corona is yet another example of a TR.

An approximate way to estimate the thermal conduction length scale for a static coronal loop can follow Eq.~(1) in \citet{john2017a}. With typical solar transition region parameters, the resulting length scale would be $\sim 30$ km or even shorter.
As pointed out by \citet{brad2013}, when the grid cell size is larger than this length scale, for a typical energetic event in the
solar atmosphere, the downward enthalpy flux cannot be handled correctly.
Part of the energy might then skip the TR and be transported directly down to the chromosphere where plasma conditions are much cooler and denser.
The consequence is that part of this energy reaches too low in the atmosphere and is radiated away by the strong optically thin radiation in the chromosphere. This in turn implies that the evaporation from the upper chromosphere or TR cannot be triggered effectively, which will finally affect the density and temperature in the solar corona.
\citet{brad2013} performed a series of one-dimensional (1D) hydrodynamic simulations with different grid cell sizes to demonstrate this problem. 
Their result showed that the coronal density of the most refined solution (reaching a grid size of $\sim 100$ m) is several times higher than that of the coarsest solution (with grid cell sizes of $\sim 400$ km), indicating that in order to fully resolve the thermal conduction in the TR, a length scale of about 1 km or even 100 m is required.

 Considering the current achievable computational resources  in the context of modern solar physics, a grid cell of about 100 m is only possible for long-term 1D simulations, assuming the length of the loop is typically 10 Mm, even with the help of parallel computing and adaptive mesh refinement (AMR).  Depending on the size of the simulation domain, a grid cell of about 1 km is very difficult or almost impossible for long-term simulations that follow dynamics in large coronal volumes for two-dimensional (2D) or three-dimensional (3D) simulations.
An approximate method is therefore required to compensate or correct for the underestimated evaporation process.

Several approaches have been proposed.
\citet{john2017a, john2017b} treated the unresolved TR as a discontinuity.
As mentioned, the evaporation is normally underestimated.
This means that the velocity at the top of the unresolved TR is incorrect.
To correct this velocity, a jump condition is derived from the conservation of energy throughout this discontinuity.
This jump condition is solved approximately to obtain a corrected velocity at the top of the unresolved TR.
In this way, the underestimated enthalpy flux is fixed.
In their tests, the results of this artificial fixing strategy agree well with their benchmark test, which was performed with a highly refined grid.
However, the accuracy of this method is limited and requires a certain range of the grid cell size, typically 100--200 km.
Moreover, to solve the nonlinear jump condition, a Newton-Raphson or similar iterator is necessary, which might affect the stability and efficiency of this method.

In contrast, another method proposed by \citet{link2001} that was later developed and implemented by \citet{lion2009} and \citet{miki2013}, appears to be more robust.
This heuristic approach introduces a fixed cutoff temperature ($T_c$) and then modifies the temperature dependence of thermal conduction and radiative cooling terms where the temperature is lower than $T_c$.
The TR would then be artificially broadened, which has the advantage that it can be numerically resolved.
This method can also help to compensate for the missing enthalpy flux, either for a static loop \citep{lion2009} or for a dynamically evolving loop \citep{miki2013}.
In addition, this method could be easy implemented not only for a 1D system, but also for multidimensional systems.
However, in some cases, especially for some strong impulsive heating problems, this method is likely to fail.
This problem comes from the use of a fixed $T_c$, which cannot always capture the correct position of the TR in such a quickly evolving system.

More recently, \citet{john2019b} combined ideas from the above two methods.
This new method is called the transition region adaptive conduction (TRAC) method.
In the TRAC method, a cutoff temperature $T_c$ is defined as well, but his time, $T_c$ is no longer a fixed value. It is calculated every time step, using the criterion defined in their previous works (e.g., \citet{john2017a, john2017b}).
Then, the authors adopted the same heuristic 
 modification for thermal conduction and radiative cooling terms as in \citet{lion2009, miki2013}.
This TRAC method is expected to be both efficient and accurate, eliminating the shortcomings of the previous two methods.
\citet{john2020} gave a detailed explanation and demonstration of the physical idea behind this method.
Based on this, an additional modification to the heating term was proposed.
At the same time, a small improvement was made to remove some abnormal jumps in the cutoff temperature between different time steps.

One topic intimately related to the evaporation process that the TRAC approach wishes to improve is the chromospheric evaporation--coronal condensation scenario \citep{anti2000}.
The evaporation will trigger thermal nonequilibrium (TNE) in the magnetic loop, which can explain the omnipresent coronal rain or prominence formation \citep{klim2019, from2020, anto2020}.
Succesful condensations also involve thermal instability (TI), a well-known linear instability \citep{park1953, fiel1965} that affects the otherwise marginal entropy mode \citep{clae2019, clae2020}.
In the evaporation-condensation scenario, a physically correct evaporation is crucial to either TNE cycles and/or to encounter conditions ripe for a condensation into solar prominence or rain due to in situ linear TI.
If the TR cannot be resolved correctly, it is possible that numerical simulations cannot produce a correct TNE cycle or prominence formation, as shown by \citet{john2019a}.

The TRAC method has so far been implemented and used only in 1D hydrodynamic simulations.
Here, we extend and apply the basic TRAC idea in 2D or 3D MHD simulations that already allow dynamic AMR.
We show a 2D simulation of prominence formation by applying two new variants of the TRAC method, which we call TRACL and TRACB.
Both of them can be extended directly to 3D simulations.
The new 2D numerical model as well as a review of the 1D TRAC method is presented in Sect. \ref{sec2}.
The result of applying our new TRACB or TRACL methods, which are directly suitable for multidimensional MHD simulations, is addressed in Sect. \ref{sec3} in representative 2D AMR-MHD runs.
Discussions and conclusions are made in Sect. \ref{sec4}.

\section{2D numerical model and 1D TRAC test}
\label{sec2}
\subsection{Simulation setup}
\begin{figure*}[h]
  \centering
  \includegraphics[width=\linewidth]{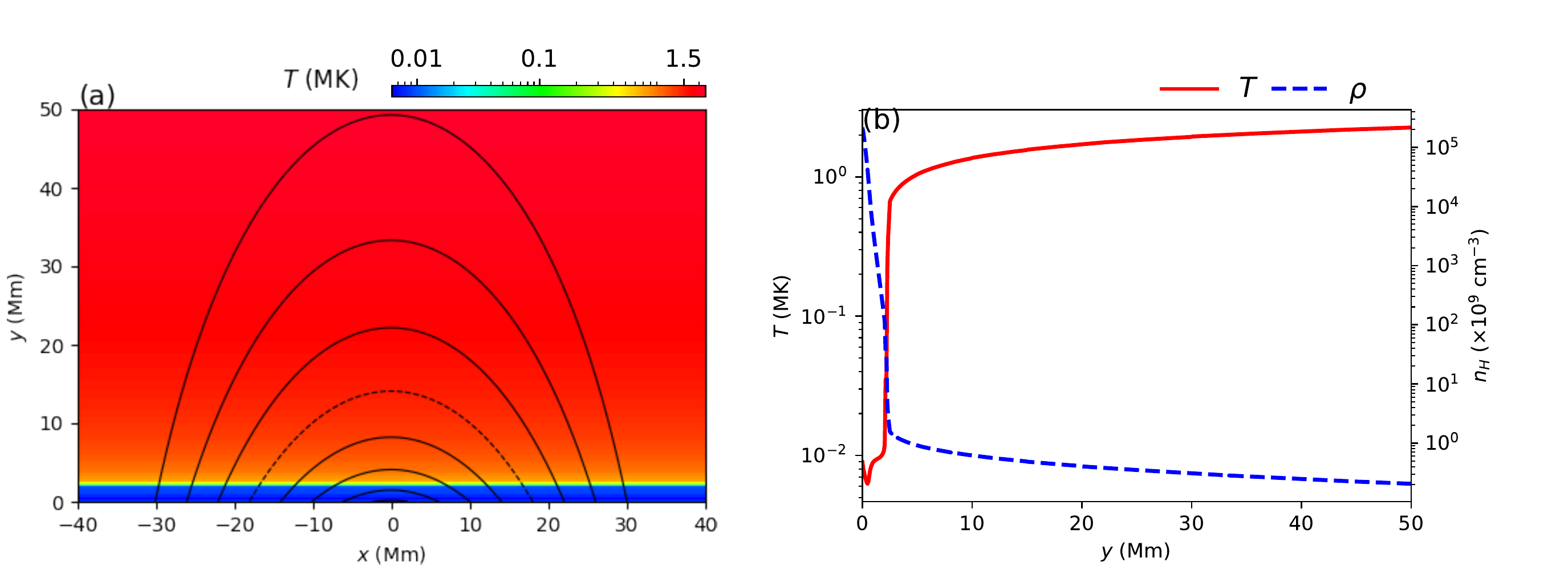}
  \caption{Initial condition of the simulation. (a) Setup of the atmosphere of our 2D simulation. Black lines show the magnetic configuration, the dashed field line is used in the 1D setup in Fig.~\ref{fig2}. (b) Initial temperature and number density profile.}
  \label{fig1}
\end{figure*}

As shown in Figure \ref{fig1}(a), the computational domain is $x_{len} = 80$ Mm in width and $y_{len} = 50$ Mm in height along the $y$-direction.
The initial force-free magnetic field is given by
\begin{equation}
  {B_x} =  - {B_0}\cos \left( {kx} \right){e^{ - j}}\cos \left( \theta  \right),
  \label{eq1}
\end{equation}
\begin{equation}
  {B_y} = {B_0}\sin \left( {kx} \right){e^{ - j}},
  \label{eq2}
\end{equation}
\begin{equation}
  {B_z} =  - {B_0}\cos \left( {kx} \right){e^{ - j}}\sin \left( \theta  \right),
  \label{eq3}
\end{equation}
where ${B_0}$ is 30 G, $k = \pi /{x_{len}}$, $\theta = \pi / 3$ and $j = k\cos \left( \theta  \right)$.
Black lines in Fig.\ref{fig1}(a) show the configuration of this magnetic field.
We used the temperature profile obtained by the 1D RADYN code \citep{hong2017}, which is relaxed from the well-known VAL-C model \citep{vern1981}, as our initial temperature distribution.
Then, given the number density $4.6 \times 10^{14}$ cm$^{-3}$ at the bottom $y=0$ boundary, the whole atmosphere can be established according to the hydrostatic equilibrium condition.
This initial temperature and density profile is shown in Figure \ref{fig1}(b).

The equations in our simulation are as follows:
\begin{equation}
\frac{{\partial \rho }}{{\partial t}} + \nabla  \cdot \left( {\rho \mathbf{v}} \right) = 0,
\label{eq4}
\end{equation}
\begin{equation}
\frac{{\partial \left( {\rho \mathbf{v}} \right)}}{{\partial t}} + \nabla  \cdot \left( {\rho \mathbf{v}\mathbf{v} + {p_{tot}}{\rm I} - \mathbf{B}\mathbf{B}} \right) = \rho \mathbf{g},
\label{eq5}
\end{equation}
\begin{equation}
\frac{{\partial e}}{{\partial t}} + \nabla  \cdot \left( {e\mathbf{v} + {p_{tot}}\mathbf{v} - \mathbf{B}\mathbf{B} \cdot \mathbf{v}} \right) = \rho \mathbf{g} \cdot \mathbf{v} + \nabla  \cdot \left( {\kappa  \cdot \nabla T} \right) - {n_e}{n_H}\Lambda \left( T \right) + H,
\label{eq6}
\end{equation}
\begin{equation}
\frac{{\partial \mathbf{B}}}{{\partial t}} + \nabla  \cdot \left( {\mathbf{v}\mathbf{B} - \mathbf{B}\mathbf{v}} \right) = \mathbf{0}.
\label{eq7}
\end{equation}
All the symbols have their usual meaning: density $\rho$, velocity $\mathbf{v}$, magnetic field $\mathbf{B}$, total pressure $p_{tot}=p+B^2/2,$ and temperature $T$.
We assumed that the whole atmosphere is fully ionized and that the ratio between the number density of helium and hydrogen, $n_{He}$/$n_{H}$, is 0.1.
The strength of the gravitational acceleration $g$ is a constant 274 m/s$^2$, and gravity acts downward.
On the right-hand side of equation (\ref{eq6}), the second term is heat conduction, where we adopted the Spitzer-type purely field-aligned heat conductivity $\kappa _\parallel = 10^{-6} T^{5/2}$ erg cm$^{-1}$ s$^{-1}$ K$^{-1}$.
The third term is optically thin radiative cooling, where the cooling table $\Lambda(T)$ is interpolated from the data given in \citet{colg2008}.
The last term $H$ is the assumed artificial heating.
Following our previous 2D simulation works on prominence formation \citep{xia2012, kepp2014, zhou2020}, this term is composed of two parts: $H_{bgr}$ and $H_{loc}$.
$H_{bgr}$, the background heating, is used to maintain the coronal temperature,
\begin{equation}
H_{bgr} = H_0\exp(-y/l_0),
\label{eq8}
\end{equation}
where $H_0=10^{-4}~\unit{erg}\unit{cm^{-3}}\unit{s^{-1}}$ and $l_0 = 50$ Mm \citep{with1977, with1988},
while $H_{loc}$ is the localized heating, used to trigger strong evaporation to induce the condensations,
\begin{equation}
{H_{loc}} = {H_1}R\left( t \right)C\left( y \right)\left[ {\exp \left( { - \frac{{{{\left( {x - {x_r}} \right)}^2}}}{{{\sigma ^2}}}} \right) + \exp \left( { - \frac{{{{\left( {x - {x_l}} \right)}^2}}}{{{\sigma ^2}}}} \right)} \right],
\label{eq9}
\end{equation}
\begin{equation}
R\left( t \right) = 
\left\{ 
{\begin{array}{*{20}{c}}
0 & \text{if } t \leq t_{relax}, \\
{\left( {t - {t_{relax}}} \right)/{t_{ramp}}} & \text{if } t_{relax} < t < t_{relax}+t_{ramp},\\
1 & \text{if } t \geq t_{relax}+t_{ramp},
\end{array}} 
\right.
\label{eq10}
\end{equation}
\begin{equation}
C\left( y \right) = 
\left\{ 
{\begin{array}{*{20}{c}}
1 & \text{if } y \leq y_h, \\
{\exp \left( { - {{\left( {y - {y_h}} \right)}^2}/{\lambda _h}^2} \right)} & \text{if } y > y_h.
\end{array}} 
\right.
\label{eq11}
\end{equation}
We adopted $H_1=2 \times 10 ^ {-2}~\unit{erg}\unit{cm^{-3}}\unit{s^{-1}}$, which is two orders higher than the background heating \citep{with1977, asch2001}.
For all other parameters, we took $x_l = -x_r = -4.2$ Mm, $y_h = 2.5$ Mm, and $\lambda _h = \sigma = 5$ Mm.
We first relaxed the system, during which time we only had $H_{bgr}$ active, for $t_{relax} = 71.5$ min. To this thermodynamically adjusted equilibrium configuration, we then gradually added $H_{loc}$ within $t_{ramp} = 500$ s.

The simulations were made with our block-based numerical code {\tt MPI-AMRVAC} \citep{kepp2012, port2014, xia2018}.
For the base level, we used 128 grid cells in the $x$ direction and 80 grid cells in the $y$ direction. Allowing a
four-level AMR leads to an effective resolution of 78 km.
A three-step Runge-Kutta time discretization \citep{shu1988} and HLL scheme \citep{hart1983} with the second-order van Leer limiter \citep{vanl1974} was used to solve equations (\ref{eq5})--(\ref{eq8}).
The STS method was used to solve the thermal conduction term, discretized as explained in~\citet{xia2018}, and the exact integration scheme was applied to solve the radiative cooling term, as mentioned in Section \ref{sec1}.
The background field splitting method in combination with an eight-wave method \citep{powe1999} were used to treat the magnetic field and control any numerical magnetic monopole errors.

For the boundary conditions, all the physics quantities were fixed at the bottom boundary while the top boundary was open.
Reflective boundary conditions were imposed on the two-side boundaries.

\subsection{1D TRAC method and test}
Before we applied the TRAC method to our 2D simulation, we briefly reviewed the original 1D version.
We tested this 1D version in our code to have an overall impression of the difference in our setup with and without the TRAC method.

The 1D TRAC method was used to treat 1D hydrodynamic equations,
\begin{equation}
\frac{{\partial \rho }}{{\partial t}} + \frac{\partial }{{\partial s}}(\rho v) = 0,
\label{eq12}
\end{equation}
\begin{equation}
\frac{\partial }{{\partial t}}(\rho v) + \frac{\partial }{{\partial s}}(\rho {v^2} + p) = \rho {g_\parallel}(s),
\label{eq13}
\end{equation}
\begin{equation}
\frac{{\partial E}}{{\partial t}} + \frac{\partial }{{\partial s}}(Ev + pv) = \rho {g_\parallel}v   + \frac{\partial }{{\partial s}}(\kappa \frac{{\partial T}}{{\partial s}})- {n_e}{n_H}\Lambda (T)+ H,
\label{eq14}
\end{equation}
where $s$ is the 1D coordinate and $g_\parallel$ is the field-aligned gravity. These are the hydrodynamic variants of our full MHD system, where $E=\rho v^2/2+p/(\gamma-1)$, in contrast to the total energy density $e=E+B^2/2$ from Eq.~(\ref{eq6}). The ratio of specific heats is $\gamma=5/3$.

According to \citet{john2019b} and \citet{john2020}, we defined the temperature length scale $L_T$ as
\begin{equation}
{L_T}\left( s \right) = \frac{T}{{\left| {dT/ds} \right|}}.
\label{eq15}
\end{equation}
When we set $L_R$ to be the size of the local grid cell (in our grid-adaptive runs, this is the smallest cell size in use in the grid hierarchy), for example,
\begin{equation}
{L_R}\left( s \right) = \Delta s \,,
\label{eq16}
\end{equation}
we can define a cutoff temperature $T_c$ as
\begin{equation}
{T_c} = \max \left( {T\left( s \right)} \right) \text{ where } \frac{{{L_R}\left( s \right)}}{{{L_T}\left( s \right)}} > \delta.
\label{eq17}
\end{equation}
Here, $\delta$ is a free parameter that should be smaller than $1/2$.
Because we used four grid points, that is, a fourth-order center difference method, to calculate the gradient of the temperature, our $\delta$ was chosen as $1/4$.
To restrict this cutoff temperature to a reasonable range, an upper bound, $0.2 \max \left( {T\left( s \right)} \right)$, and a lower bound, $2 \times 10^4$ K, were set, as recommended by \citet{john2019b}.
Other restrictions to prevent some sudden jumps can be found in the Appendix A of \citet{john2020}.

For the region in which the temperature is lower than $T_c$, the TRAC method applies the following modifications: 
\begin{equation}
\kappa'  = \kappa T_c^{5/2} \,,
\label{eq18}
\end{equation}
\begin{equation}
\Lambda '\left( T \right) = \Lambda \left( T \right){\left( {\frac{T}{{{T_c}}}} \right)^{5/2}}\,,
\label{eq19}
\end{equation}
\begin{equation}
H' = H{\left( {\frac{T}{{{T_c}}}} \right)^{5/2}}\,.
\label{eq20}
\end{equation}
The left-hand side symbols with prime are used to substitute the original quantities in equation~(\ref{eq14}).

\begin{figure*}[h]
  \centering
  \includegraphics[width=0.8\linewidth]{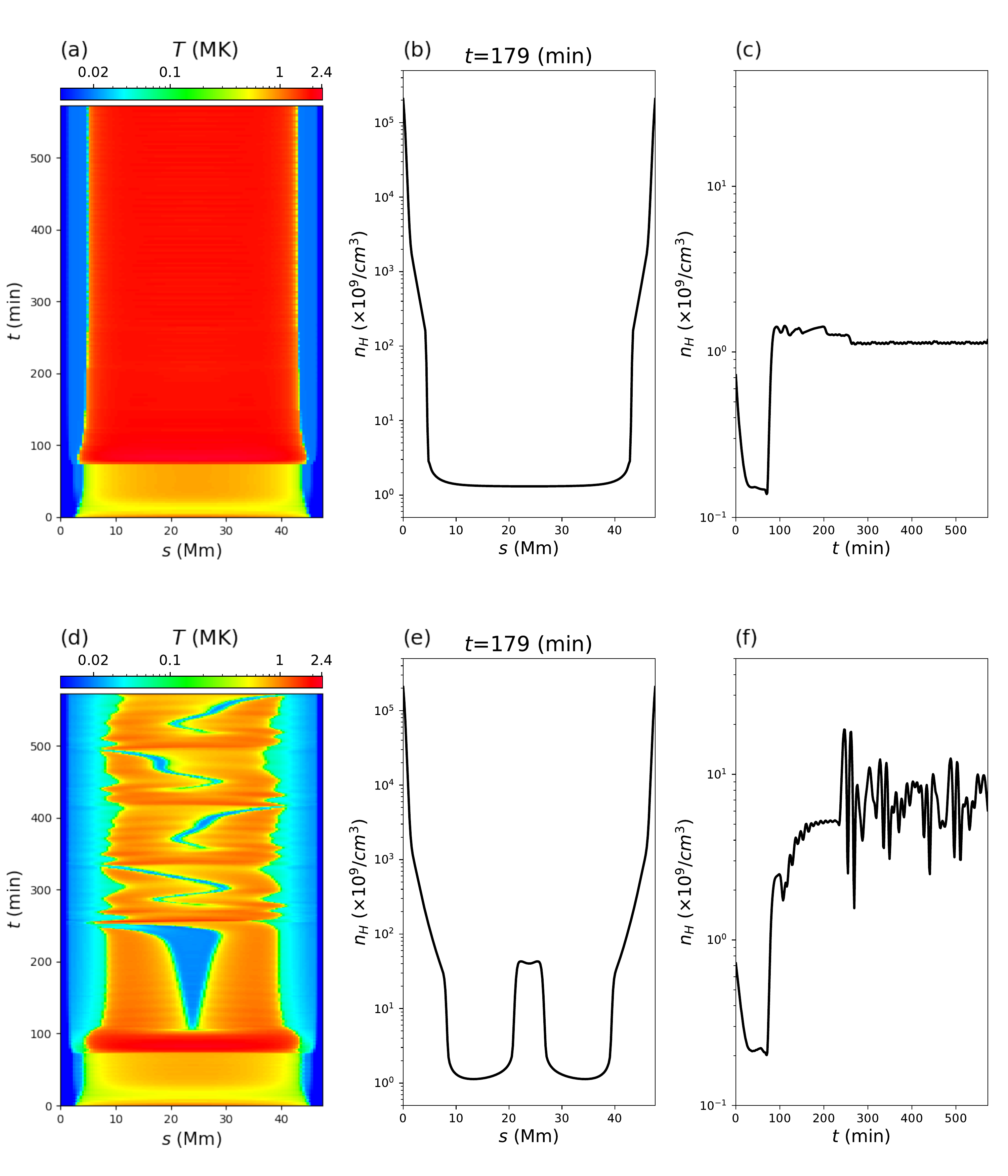}
  \caption{Simulation results of the 1D tests. Left column: Temperature evolution of the 1D loop extracted from Fig.~\ref{fig1}. Middle column: Number density distribution along the loop at $t = 179$ min. Right column: Evolution of the averaged coronal density. The upper panels show results without the TRAC method, and the lower panels show results with the TRAC method.}
  \label{fig2}
\end{figure*}

In order to test the 1D TRAC method, we picked out a field line at $t=0$ in our 2D setup, which is then a coronal loop, starting from point $x = -18$ Mm at the bottom boundary. This field line is shown by the dashed line in Fig.~\ref{fig1}(a). The total length of this loop is 47 Mm.
We now discretized it with 160 grid cells at the base level, using a
three-level AMR, leading to an effective resolution of 74 Mm, which is similar to the size of our 2D grid cells (78 Mm).
All the other settings were adopted directly from the 2D setting described above.
Assuming that the shape of this magnetic loop remains invariant, we now ran the equivalent 1D hydro-simulation for over 9 h. The result is shown in Fig.~\ref{fig2}.

The upper panels show the result of this 1D hydro-run without the TRAC method, and the lower three panels show the results with the TRAC method.
Panels (a) and (d) show the temperature evolutions for both runs.
When no TRAC method is used, only a hot coronal loop with gradually increasing temperature is visible; the increase occurs very slowly.
However, when the TRAC method is applied, thermal instability occurs after we add the localized heating
for about half an hour, so that condensation sets in: a prominence therefore forms in this case.
Panels (b) and (e) show the number density distribution along the loop at $t=179$ min.
The density of the prominence is nearly two orders higher than the coronal density.

In our previous works \citep{xia2011, zhou2014}, this type of 1D simulations was performed in a dipped flux tube configuration, so that the formed condensation was gravitationally trapped and remained very stable for days or even weeks. In the 2D interpretation of such a rigid magnetic field 1D run, the dip corresponds to an upward Lorentz force.
In the situation of Fig.~\ref{fig2}, however, the configuration is a simple arcade, which cannot provide upward support.
Because the flux tube is treated as a rigid body in a 1D simulation,
gravity from the heavy prominence cannot change the shape of the prescribed flux tube.
The prominence therefore eventually slides down to the chromosphere.
This occurs at about $t=250$ min.
Afterward, evaporation and condensation continuously take place, forming TNE cycles.
These results are very similar to those of \citet{john2019a}.

Panels (c) and (f) show the evolution of the averaged coronal number density for both runs.
Here, the corona is simply defined as the region between $s=13$ Mm to $s=34$ Mm.
Clearly, the TRAC method improves the evaporation and results in a higher coronal density.
The averaged density is typically several times the density found without the TRAC method.

Only in a higher coronal density situation can the thermal instability (this hinges on the quadratic density dependence of the optically thin radiative loss term) as well as the condensation occur.
Therefore we conclude that for this type of prominence formation simulation, the TRAC method is very important.
This is all clearly very sensitive to the employed heating prescription, but when the TRAC method is used, we can meaningfully explore a relatively large parameter space, beyond what is possible without using TRAC, at much more affordable grid resolutions.

\section{Applying the TRAC method in the 2D setup}
\label{sec3}
Because thermal conduction in multidimensional MHD is aligned with the possibly relocating field lines, the key point of applying a TRAC method in a 2D MHD simulation is the requirement of tracing magnetic field lines from time step to time step.
We have recently \citep{ruan2020} developed a magnetic field-line-tracing module for {\tt MPI-AMRVAC} to solve the fast electron heating problem in the context of solar flares, where field-aligned fast electron beams deposit their energy in the denser atmospheric regions after gaining energy in a coronal reconnection site.
Based on the experience there, we now propose two different approaches for applying the TRAC idea in multidimensional MHD settings.
While details are provided in the appendix, we briefly introduce the main ideasin the following subsections.

\subsection{TRACL}
A straightforward field-line based or TRACL method is directly using the magnetic field-tracing module from our previous work.
For this simulation, we have $128\times80$ grid cells with four AMR levels, which means that we effectively have $1024\times640$ grid cells.
Because the arcade configuration is not expected to undergo dramatic changes, we chose to trace 256 field lines whose starting points are uniformly distributed on the lower $y=0$ boundary. 
By tracing the field lines at every time step, we can calculate the cutoff temperature $T_c$ for each of them. This $T_c$ then essentially changes from field line to field line, and it also changes with time.
Then, by interpolating $T_c$ from the field lines back onto the AMR grid cells, we obtain the distribution of the cutoff temperature for all grid cells.
Consequently, we can applied equations~(\ref{eq18})--(\ref{eq20}) to modify the local evaluations of these thermal terms, similar to the previous 1D situation.
This TRAC method is based on directly integrating field lines, and is appropriately named TRACL.

\begin{figure*}[h]
  \centering
  \includegraphics[width=0.8\linewidth]{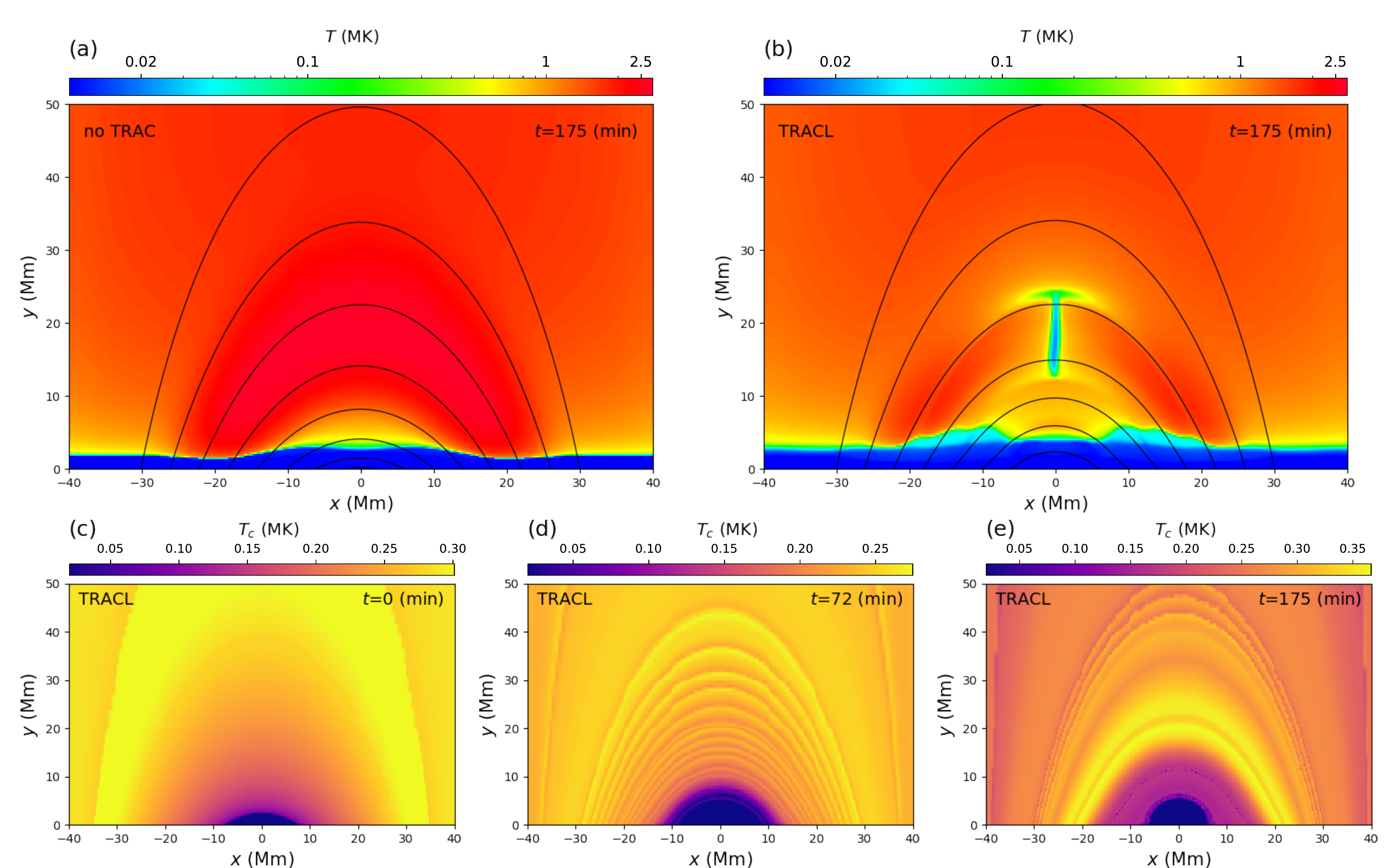}
  \caption{Simulation results of the TRACL method. Upper panels: Temperature distributions at $t=175$ min for (a) the case without TRAC and (b) with TRACL. Lower panels: $T_c$ distributions for the TRACL case at different times: initially (c), after relaxation, and (d) at the end (e).}
  \label{fig3}
\end{figure*}

To demonstrate its effect, we ran two cases, one without any TRAC method and another with this TRACL method.
The results are shown in Fig.~\ref{fig3}.
Panels (a) and (b) are snapshots at $t=175$ min for the two different cases.
Similarly to what happened in the 1D case, when we did not use the TRACL method, only a hot loop develops in panel (a). This shows that without TRACL, there is still no condensation even at the end of our simulation after $t=250$ min.
In panel (b) the same setting with the TRACL method is shown. The formation of a solar prominence whose typical lower temperature is $2\times10^4$ K is clearly visble, corresponding to the lower boundary of our employed cooling table.

The distributions of the cutoff temperature $T_c$ at different times are shown in the lower row of Fig. ~\ref{fig3}.
Panel (c) shows its distribution at $t=0$.
The lower and shorter loops have a lower cutoff temperature, and the higher and longer loops tend to have a higher cutoff temperature.
The shapes of these initial cutoff temperature contours are clearly consistent with the magnetic field lines, which shows that our magnetic field-line-tracing method is robust.
After relaxation at $t=72$ minutes, the pattern adjusts, as shown in Fig.~\ref{fig3}, panel~(d).
The lower cutoff temperature region near the center has grown: although the temperature in this region increases, the resulting cutoff temperature decreases.
In the last panel (e), after some time of evaporation, we show the cutoff temperature distribution  at $t=175$ minutes, when condensation has begun.
The cutoff temperature of the field lines, which went through the heated loop that then collapsed into a filament, is slightly higher than in the outside loops. The field lines have (slightly) adjusted during the entire run, and especially show some minor central dipped portion that develops at this end time.

\subsection{TRACB}
From the above numerical experiment, we conclude that the TRACL method can successfully correct for underestimated evaporation.
Nevertheless, integrating many magnetic field lines at every time step, the core of the TRACL method, is a time-consuming process. In a modern parallelized simulation, global communication is required every time the magnetic field line goes across processors, and this happens several times for each field line \citep{ruan2020}.
Considering that the precise value of $T_c$ is not necessarily highly accurate, we here advocate an approximate and fast way to trace the magnetic field lines in setups such as the one studied here.

Our {\tt MPI-AMRVAC} code is a block-based AMR code, which means that the computational domain is first discretized into blocks, and these blocks all have an equal amount of grid cells.
Each block is composed of several grid cells at the same AMR level, and
for this 2D simulation, we chose the block size to be $16\times16$ ({\tt MPI-AMRVAC} allows changing this setting at runtime).
Our effective resolution of $1024\times640$ grid cells means that we handle $64\times40$ blocks at most. We may therefore 
calculate the local cutoff temperature $T_{c, local}$ using equation (\ref{eq17}) for each block. 
Then, the maximum operator in equation (\ref{eq17}) is first done locally in each block, and only this
$T_{c, local}$, as well as the unit vector of the magnetic field at the center of this block, denoted as $\bm{b}_{local}$, is then broadcasted to all the processors.
By sharing this $64\times40$ global array consisting of magnetic field unit vectors and cutoff temperatures, the field line can be traced independently in each processor using this information.
Again, the cutoff temperature along the (coarsely represented) field lines is then interpolated back onto the AMR grid cells, so that we have the distribution of $T_c$ throughout the whole computational domain.
Because this method is based on blocks, we call it the TRACB method.
Additional global communication is required only once for each time step, so tat the computational efficiency is expected to be higher than for the TRACL method, at the cost of a coarser evaluation of the cutoff temperature distribution.

\begin{figure*}
  \centering
  \includegraphics[width=0.8\linewidth]{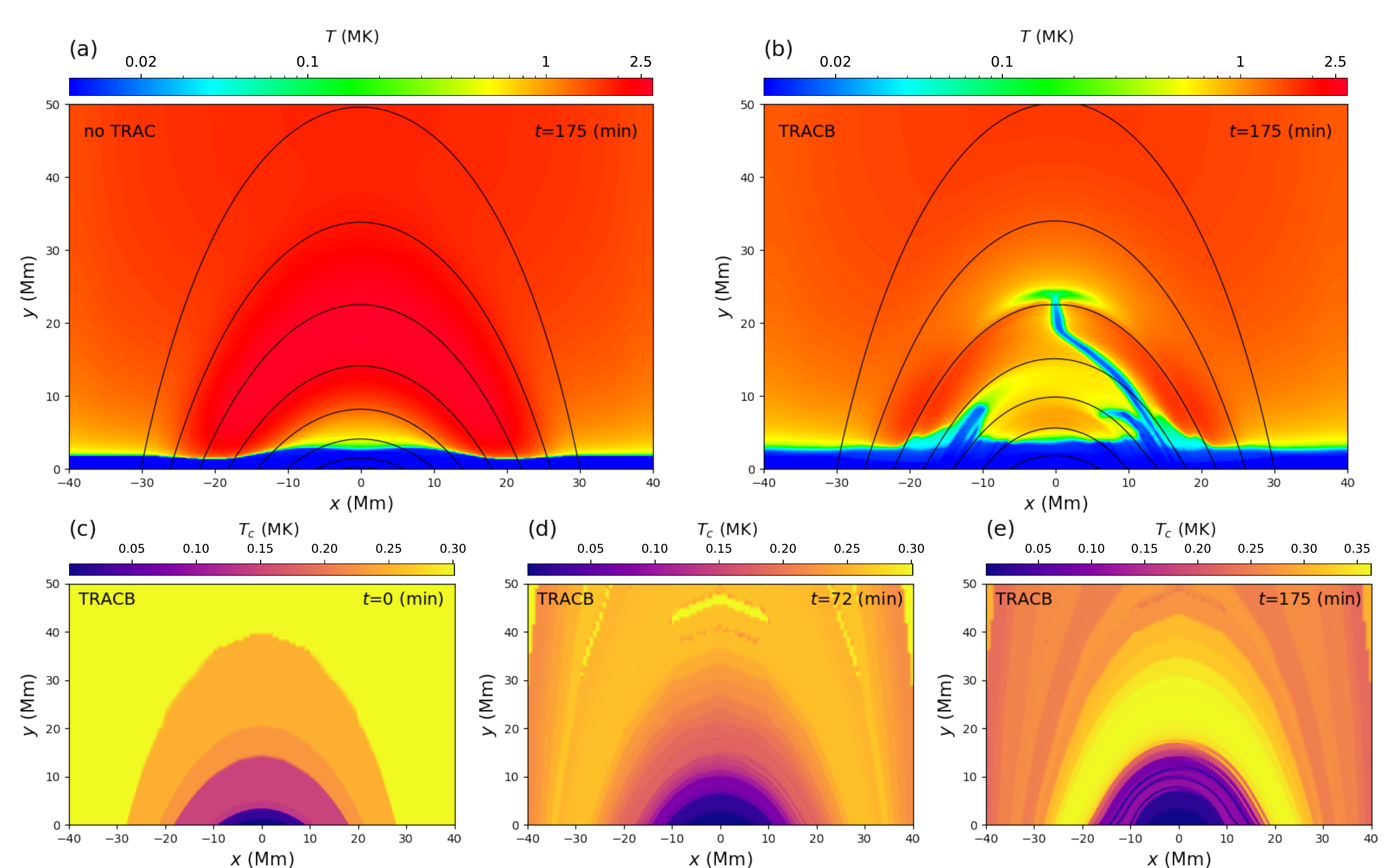}
  \caption{Similar to Figure~\ref{fig3}, but this time showing results for the TRACB case in panels (b), (c), (d), and (e).}
  \label{fig4}
\end{figure*}

We ran the same simulation as in the previous subsection, but applied the TRACB method instead.
This result is shown in Fig.~\ref{fig4}: as in Fig.~\ref{fig3}, panel (a) is the result without the TRAC method, for comparison.
With TRACB, panel (b) shows that the condensation indeed occurs, similar to the TRACL case.
The atmosphere is similar to that in Fig.~\ref{fig3}(b), and the condensation occurs at the same position, as expected.
However, because this method only employs a fairly coarse cutoff temperature distribution, we find that this time, the prominence is not as stable, and especially the left-right symmetry is easily broken.
The newly formed prominence therefore has already started to slide down to one side, similar to what we obtained in the 1D simulation, for instance, after $t=250$ min in Fig.~\ref{fig2}(d).

Again, in the lower row of Fig.~\ref{fig4}, we show the distribution of $T_c$ at $t=0$, 72, and 275 minutes.
For the initial snapshot in panel (c), the distribution is similar to Fig.~\ref{fig3}, but clearly not as smoothly represented as with the TRACL method.
The cutoff distribution is only aware of magnetic field lines as traced by the block-center magnetic field. For this initial state, only using the block center magnetic field can also trace the magnetic field lines and give a reliable result. However, after relaxation, as shown in panel (d), the result differs slightly from the TRACL method.
This is mainly because equation~(\ref{eq17}) relies on a constant $\delta$ parameter.
Such a fixed value throughout the whole domain, in combination with the hierarchical AMR grid structure, will sometimes result in a hierarchical $T_c$ structure, as shown in Fig. \ref{fig3}(d), which might be considered a drawback of the TRACB method.
In the TRACB method, the final distribution of $T_c$ relies more heavily on interpolation.
Then, for the final panel at the time $t=175$ minutes, the distribution is again similar with Fig. \ref{fig3}(e), especially when we recognize that the atmosphere itself is already slightly different.
In the last two panels (d) and (e), some regions near the upper corner are not appropriately covered by the field lines we selected because the AMR structure uses a coarser grid there. In most applications, the cutoff temperature at the upper corners does not affect the result, we simply filled them with some arbitrary values.

\subsection{Comparing TRACL and TRACB}
\begin{figure*}
  \centering
  \includegraphics[width=0.8\linewidth]{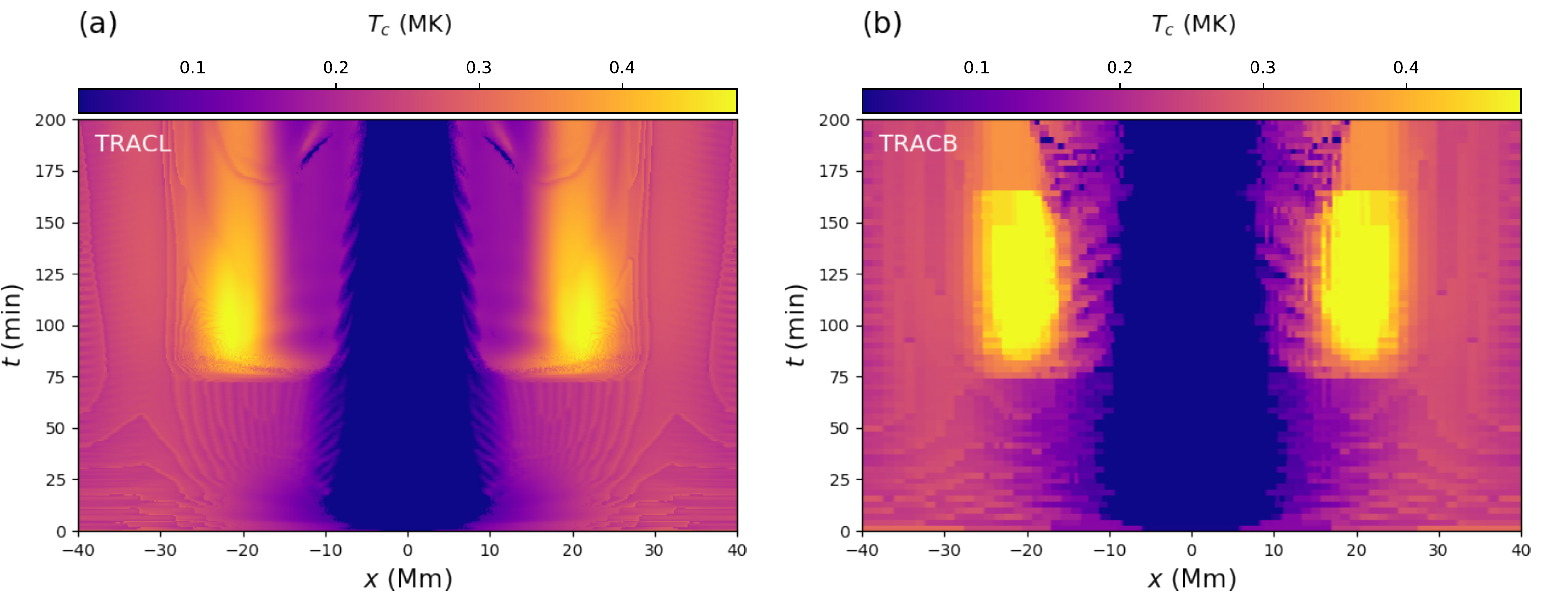}
  \caption{Time evolution of $T_c$ distributed along the $y=2$ Mm line for (a) the TRACL case and (b) the TRACB case.}
  \label{fig5}
\end{figure*}

To be more precise in the comparison of the two new TRAC methods, we cut a slice along the horizontal line $y=2$ Mm and quantified the time evolution of $T_c$.
The result is shown in Fig.~\ref{fig5}.
The left panel shows the TRACL and the right panel the TRACB method.
The short loops near the center always have the lowest temperature, which means that along these corresponding field lines, it is not necessary to modify the TRAC from the criterion expressed in equation~(\ref{eq17}).
However, on either side of this central region, the typical $T_c$ is about $2\times10^5$--$3\times10^5$ K for a steady loop.
After the localized heating is introduced, $T_c$ near the heating source increases quickly and maintains a high temperature value of over $4\times10^5$ K.
Both methods show a similar pattern in the temporal evolution of the cutoff temperature at this height, although there are noticable differences, mostly confirming that TRACB yields a more coarser representation of the actual evolution.

\section{Discussion and conclusions}
\label{sec4}

\subsection{From 1D hydro- to multidimensional MHD}
Although the TR is a very thin layer in the solar atmosphere, it plays a crucial role in many phenomena occurring in the solar atmosphere.
Traditionally, the TR is considered to be the region in which temperature is about 10$^5$ K, between the chromosphere and the corona. Many of our own previous numerical simulations that related to TR evaporation, such as prominence formation \citep{xia2012, kepp2014, xia2016} or coronal rain formation \citep{fang2013, fang2015, xia2017}, did not fully resolve the TR, so that the evaporation was underestimated. The combination of modern shock-capturing schemes, employing limited linear or higher order reconstructions in our finite volume code, did not preclude a numerically stable representation of TRs, but this at the expense of an altered energy balance from the true TR regime.
Then, with insufficient evaporation, the density in the solar corona would be severely affected, as shown by \citet{brad2013} in their 1D hydro-simulation.
This is mainly because the enthalpy flux in the transition region is underestimated in an unresolved TR.

For 1D problems, brute force high resolution that is for example efficiently achieved with AMR could be used to solve this problem.
However, for multidimensional problems, a similarly high resolution is beyond computational reach, especially because the requirement might be even higher than estimated from 1D simulations. The 1D case shown in Sec. \ref{sec2} had an effective resolution of 74 km, similar to the 78 km, that is, the effective resolution of the 2D case in Sec. \ref{sec3}, which used four AMR levels. In the 1D case, we find that when we add just one more grid level, that is, resolve to 37 km, the condensation starts to occur in a non-TRAC simulation, at nearly the same time as the 1D TRAC case.
However, for the 2D case, even when we increased to seven AMR levels and run the simulation for 3.5 h, we still did not see any condensation when we did not use TRACB or TRACL.

\begin{figure}
  \centering
  \includegraphics[width=\linewidth]{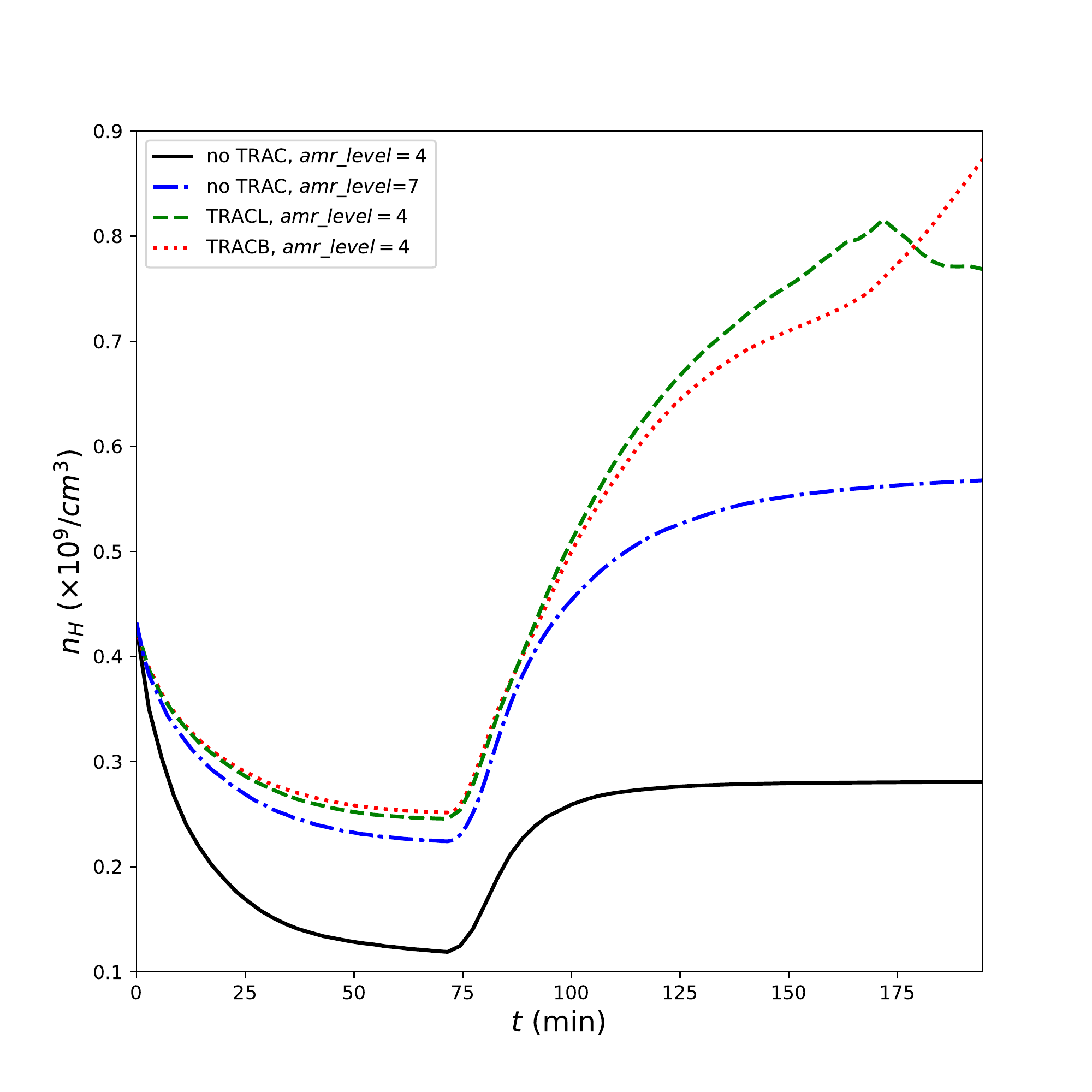}
  \caption{Time evolution of the averaged coronal density for different cases, as shown in the legend.}
  \label{fig6}
\end{figure}

There are two distinct possibilities: (a) the resolution required to correctly represent the TR is more strict in 2D simulations, or (b) the TRACL/TRACB methods in 2D give unphysical results. To judge this aspect and argue in favor of option (a), 
Fig.~\ref{fig6} shows the time-evolution averaged coronal density for different cases.
For the same number of four AMR levels, the run without the TRAC method solid line is much lower than the two runs with our new TRACB and TRACL methods dotted and dashed lines.
This is expected from the design of the TRAC method, and is meant to capture the effect of a higher resolution that is needed to obtain the proper evaporation. However, the dash-dotted line represents the average coronal density evolution for a high-resolution run (employing seven AMR levels) without any TRAC method.
Before the localized heating is added (before 75 minutes) and when the localized heating starts to work, this reference result is very close to the two dashed lines. However, after about $t=100$ minutes, it deviates from the two dashed lines, well before the condensations that occur in the last two cases (at about $t=120$ minutes).
Based on this result, we argue that multidimensional simulations require an even higher resolution when the intricacies of anisotropic thermal conduction along changing and evolving magnetic field lines are taken into account.

\subsection{Efficiency considerations}
\label{sec4.2}
The 1D TRAC method is a fast and accurate method, and in our {\tt MPI-AMRVAC} implementation, the computational time for a 1D TRAC simulation is almost the same as without the TRAC method.
Considering that the modified temperature prescription in the thermal conduction term allows us to use a larger time step in each STS substep, the TRAC method sometimes is even faster.
However, applying it in a multidimensional simulation requires additional time spent on tracing the magnetic field lines.
\citet{ruan2020} developed and used a generic magnetic field-line-tracing module for our open-source code. In this application, we did not (yet) optimize the field-line-tracing module, and hence it was very time-consuming and performed poorly when the number of processors was increased (because the many field lines were processed serially). For the results presented in this paper, we optimized this module by parallelizing it and allowing a reliable result using the TRACL method, obtained within a reasonable time span.

Considering  the block-based structure of our code, we implemented and presented an alternative block-based TRACB method, keeping the essential functionality of estimating the $T_c$ distribution over the domain.
This TRACB method is mostly performed locally in each processor, with only one additional global communication every time step. This is expected to improve the overall efficiency.

We have also experimented with yet another way to gain efficiency by noting that the TRAC method only modifies the region in which the temperature is lower than the cutoff temperature, which is typically the region below the TR.
This means that integrating the magnetic field lines throughout the whole computational domain is slightly in vain as long as no coronal condensations have formed.
In practice, we can therefore only integrate field lines up to a certain height, for example, 8 Mm or 10 Mm, and mask out the higher coronal part.
This will save memory space as well as computational resources, and although we did not show this here, the results are quite similar to the unmasked version that we used throughout her (in both TRACL and TRACB). This may be a crucial ingredient for actual 3D simulations using TRACL or TRACB variants.

In\citet{john2019b} and\citet{john2020}, the upper value of $T_c$ was set to $T_{ub}=0.2\max(T(s))$.
In our applications, we also gave this upper boundary to $T_c$ when each magnetic field line was traces.
When we choose to integrate only part of the field lines, as suggested in the previous paragraph, we cannot obtain the maximum temperature for each field line.
However, this upper bound is not so frequently invoked in the current setup.
It is therefore not necessary to distinguish $T_{ub}$ from field line to field line,
and the overall maximum temperature can instead be used over the full grid as a good substitute.
For instance, for flare-related simulations that reach truly high temperatures, setting this global $ T_{ub}$ might also be a good choice.

Table \ref{tab1} quantifies the computational time used for the simulation described in Section \ref{sec2}, as run by the different methods mentioned above. When field lines are only integrated in a region below 8 Mm, we refer to our result as TRACL$_\text{mask}$ and TRACB$_\text{mask}$. For a fair comparison, we ran the simulation always from $t=0$ to $t=71.5$ min, that is, before we added the localized heating, so that the number of blocks on different AMR levels is the same or at least very similar.
The simulations were run on two different machines, a modern desktop with 16 processors, and a Tier-1 cluster, where we employed only up to 56 processors.
For the desktop, the processor is a Xeon E5-2630 v3 with a maximum frequency of 2.4 GHz. 
On the cluster, we used two nodes, and each node had two Xeon 6132 CPUs with a highest frequency of 2.6 GHz. On both machines, the Gfortran compiler was used.

\begin{table}[!h]
\begin{tabular}{ccc}
\hline
TRAC method & desktop (min) & clusters (min) \\
\hline
no TRAC & 440 & 135 \\
TRACL & 710 & 703 \\
TRACL$_\text{mask}$ & 576 & 544 \\
TRACB & 474 & 165 \\
TRACB$_\text{mask}$ & 407 & 126 \\
\hline
\end{tabular}
\caption{Computational time for the different methods.}
\label{tab1}
\end{table}

From these timing results, we conclude that for the TRACL method, masking the coronal part may well be necessary for large-scale 2D and ultimately 3D problems.
If a full-domain integration is necessary for the problem at hand, the TRACB method could be used, while the most efficient implementation is our TRACB$_\text{mask}$ approach.

\subsection{Broadening of the TR}
As discussed in our introduction, before the TRAC method, several other methods aimed to solve the unresolved TR problem.
The method by \citet{link2001} and \citet{lion2009}, using a globally fixed cutoff temperature $T_c$, could be considered as the original idea behind the TRAC method.
An actual unresolved TR is then artificially broadened when the thermal terms are considered, and hence becomes a resolved TR. This is achieved by keeping the product of $\kappa \Lambda$ in equations (\ref{eq18})--(\ref{eq19}) unchanged.
However, a globally fixed cutoff temperature cannot always capture the correct position of the TR because, as proposed by \citet{brad2010a, brad2010b}, from the physical point of view, the interface between the TR and the corona should be the point where the enthalpy flux changes from a sink into a source.
However, this criterion is not so easily implemented into a numerical MHD code. \citet{john2017a, john2017b} therefore defined the top of the TR by an artificial correction formula, which was then used in the TRAC method; this  is our equation~(\ref{eq17}) here.

\begin{figure*}
  \centering
  \includegraphics[width=0.8\linewidth]{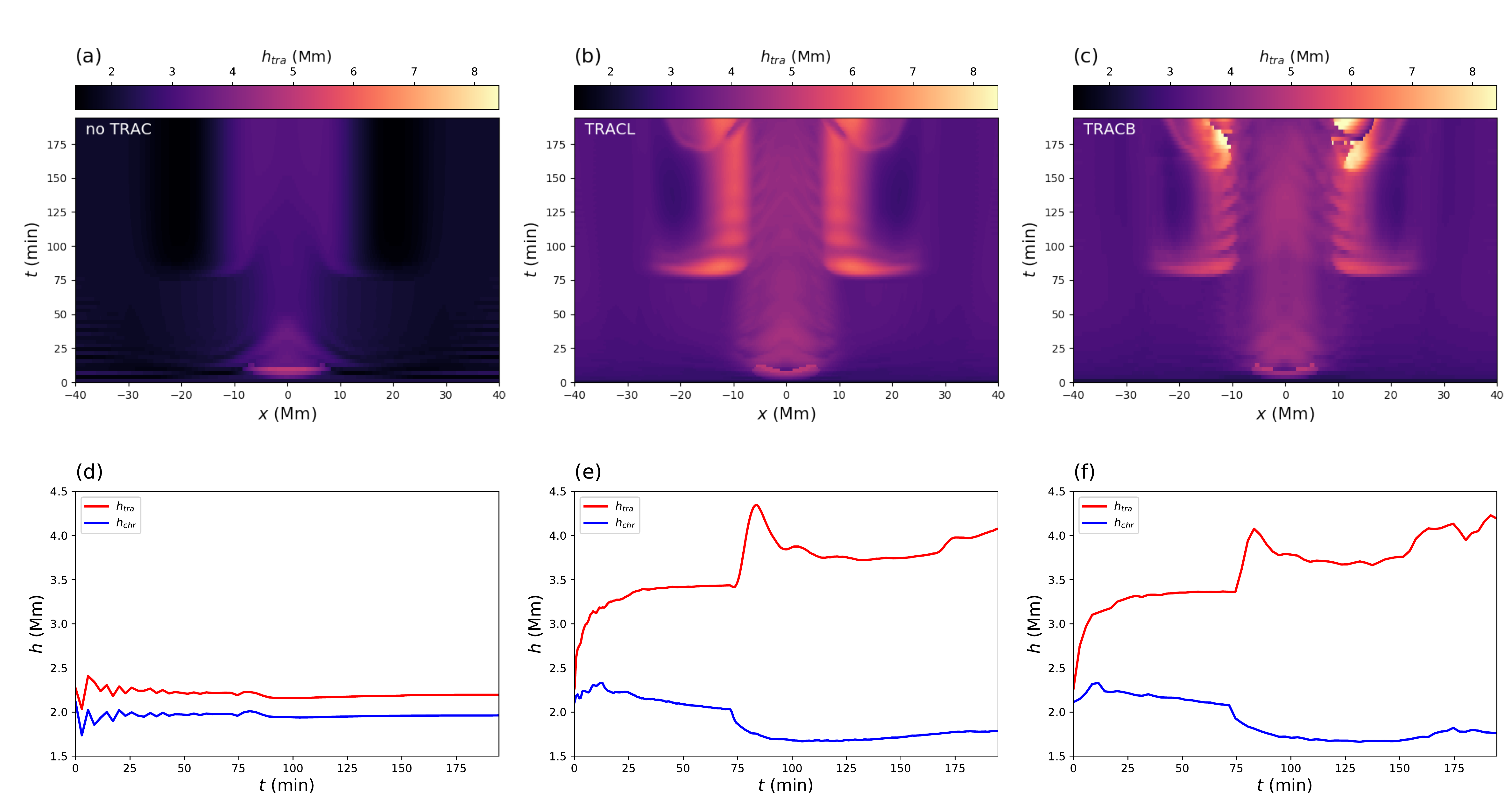}
  \caption{Time evolution of the transition region. Upper panels: Time evolution of the height of the transition region $h_{tra}$ for different runs: no TRAC (a), TRACL (b), or TRACB (c). Lower panels: Time evolution of $h_{tra}$ and $h_{chr}$ as averaged along the $x$ direction for the corresponding panels above.}
  \label{fig7}
\end{figure*}

The effect of the TRAC method on the height of the TR is shown in Fig. \ref{fig7}.
The left column shows the run without the TRAC method, the middle column shows the TRACL method, and the right column shows the result with the TRACB method.
The upper row shows the time evolution of the height of the top of the TR, $h_{tra}$.
Here, the top of the TR is simply defined as the place at which the temperature gradient along the vertical direction ($y$-direction) is maximum. Directly using
equation (\ref{eq17}) is not appropriate because for the short loops at the center, it is possible that no points on the loop fulfill this threshold.
The time evolution of the horizontally ($x$-direction) averaged transition region height $h_{tra}$ is plotted in the lower row, as well as the averaged height of the bottom of the TR denoted with $h_{chr}$. This
$h_{chr}$ is defined as the first point from the bottom at which the temperature exceeds $2\times10^4$ K.
When no TRAC is applied, the TR does not have any obvious response to the strong localized heating.
After we introduced the localized heating, $h_{tra}$ near the source region decreased slightly while the center part increased slightly, but the overall change is minor without TRAC.
In contrast, when the TRAC method was used, the TR broadened, especially after we introduced the localized heating.
The width of the TR was  four to six times broader.
The evolution of the TRACL method and the TRACB method is basically very similar, with some deviations near the end of the simulation.

In this example and in other simulations that follow the evaporation--condensation scenario, we are interested in the time and location at which condensations occur in the solar corona. With that in mind, having the correct density-temperature throughout the corona is more important than truly resolving a `correct' TR.
A broadened TR is acceptable in this type of simulations. However, when condensations do occur (directly related to TI onset), another "TR" develops between the prominence or coronal rain blob and the solar corona. This is the prominence--corona transition region (PCTR). When equation (\ref{eq17}) is used, this PCTR also broadens artificially.
The PCTR broadening is not as strong as the broadening in the TR because a much weaker heating acts at these heights. When we instead require that the PCTR not be modified at all, as mentioned in Sec. \ref{sec4.2}, we can add the mask and use TRACB$_\text{mask}$ or TRACL$_\text{mask}$ approaches.

\subsection{Conclusion}
Chromospheric evaporation is an important source of mass and energy for the solar corona.
However, the TR between the chromosphere and corona has a very steep temperature gradient, and this implies a real challenge for numerical simulations, especially when we consider the anisotropic nature of thermal conduction. Earlier 1D hydrodynamic settings have shown that using a too 
low grid resolution, that is, one that cannot resolve the TR gradients fully, might finally underestimate the evaporation, resulting in too low coronal densities. 
The TRAC method \citep{john2019b, john2020} is an ad hoc way to correct for these problems, and this method has been investigated in 1D simulations so far.

In this work, we introduced two novel means to apply the TRAC method in multidimensional simulations that employ block-grid-adaptive techniques to efficiently resolve chromosphere to coronal dynamics. Both involve the use of a dynamic field line tracing from the chromosphere to the corona, along which an essentially 1D approach is then employed. Because modern simulations use parallel computers where the computational domain is split into blocks that are spread over the CPUs, such field line tracing must handle the complexities of field line relocations and evolutions that involve multiple processors. In the most modern applications, this parallelism is combined with AMR, introducing the dynamic creation and destruction of hierarchical grid blocks. Our TRACL and TRACB methods are fully compatible with AMR, but can also easily be used in pure domain-decomposed MHD simulations that use a simple structured grid throughout. Both methods were tested here and were implemented in the open-source {\tt MPI-AMRVAC} code. They are available and integrated in our open-source software.

Although the example presented here is 2D, it is trivial to extend the method from 2D to 3D, and it is also already implemented in the dimension-independent {\tt MPI-AMRVAC} code. In 3D, the TRACB method shows a higher efficiency than the TRACL method.
The magnetic configuration in this example 2D scenario is a relatively simple arcade structure. No reconnection occurs during the times that are simulated.
However, the magnetic field-line-tracing method is robust and can be applied to more complex configurations such as reconnection events, as shown in \citet{ruan2020}, who followed a flare event.
One notable future improvement might be the choice of the initial points used to integrate field lines.
Currently, the initial points are distributed near the bottom boundary, which is applicable in most cases.
This might not be the best choice for some configurations, especially near side boundaries in a complex realistic field.

\begin{acknowledgements}
This work was supported by a joint FWO-NSFC grant G0E9619N and received funding from the European Research Council (ERC) under the European Unions Horizon 2020 research and innovation program (grant agreement No. 833251 PROMINENT ERC-ADG 2018).
W.R. received funding from the Chinese Scholarship Council. C. X. is supported by the National Natural Science Foundation of China (NSFC, Grant No. 11803031 and 12073022) and the Basic Research Program of Yunnan Province (Grant No. 2019FB140 and 202001AW070011). This research is further supported by Internal funds KU Leuven, project C14/19/089 TRACESpace. The computational resources and services used in this work were provided by the VSC (Flemish Supercomputer Center), funded by the Research Foundation Flanders (FWO) and the Flemish Government - department EWI.
\end{acknowledgements}

\bibliographystyle{aa}
\bibliography{ref}

\begin{thebibliography}{55}
\expandafter\ifx\csname natexlab\endcsname\relax\def\natexlab#1{#1}\fi

\bibitem[{Alexiades {et~al.}(1996)Alexiades, Amiez, \& Gremaud}]{alex1996}
Alexiades, V., Amiez, G., \& Gremaud, P.-A. 1996, Communications in numerical
  methods in engineering, 12, 31

\bibitem[{{Antiochos} {et~al.}(2000){Antiochos}, {MacNeice}, \&
  {Spicer}}]{anti2000}
{Antiochos}, S.~K., {MacNeice}, P.~J., \& {Spicer}, D.~S. 2000, \apj, 536, 494

\bibitem[{{Antolin}(2020)}]{anto2020}
{Antolin}, P. 2020, Plasma Physics and Controlled Fusion, 62, 014016

\bibitem[{{Aschwanden}(2001)}]{asch2001}
{Aschwanden}, M.~J. 2001, \apj, 560, 1035

\bibitem[{{Bradshaw} \& {Cargill}(2010{\natexlab{a}})}]{brad2010a}
{Bradshaw}, S.~J. \& {Cargill}, P.~J. 2010{\natexlab{a}}, \apjl, 710, L39

\bibitem[{{Bradshaw} \& {Cargill}(2010{\natexlab{b}})}]{brad2010b}
{Bradshaw}, S.~J. \& {Cargill}, P.~J. 2010{\natexlab{b}}, \apj, 717, 163

\bibitem[{{Bradshaw} \& {Cargill}(2013)}]{brad2013}
{Bradshaw}, S.~J. \& {Cargill}, P.~J. 2013, \apj, 770, 12

\bibitem[{{Cheung} {et~al.}(2019){Cheung}, {Rempel}, {Chintzoglou}, {Chen},
  {Testa}, {Mart{\'\i}nez-Sykora}, {Sainz Dalda}, {DeRosa}, {Malanushenko},
  {Hansteen}, {De Pontieu}, {Carlsson}, {Gudiksen}, \& {McIntosh}}]{cheu2019}
{Cheung}, M.~C.~M., {Rempel}, M., {Chintzoglou}, G., {et~al.} 2019, Nature
  Astronomy, 3, 160

\bibitem[{{Claes} \& {Keppens}(2019)}]{clae2019}
{Claes}, N. \& {Keppens}, R. 2019, \aap, 624, A96

\bibitem[{{Claes} {et~al.}(2020){Claes}, {Keppens}, \& {Xia}}]{clae2020}
{Claes}, N., {Keppens}, R., \& {Xia}, C. 2020, \aap, 636, A112

\bibitem[{{Colgan} {et~al.}(2008){Colgan}, {Abdallah}, {Sherrill}, {Foster},
  {Fontes}, \& {Feldman}}]{colg2008}
{Colgan}, J., {Abdallah}, Jr., J., {Sherrill}, M.~E., {et~al.} 2008, \apj, 689,
  585

\bibitem[{{Fang} {et~al.}(2013){Fang}, {Xia}, \& {Keppens}}]{fang2013}
{Fang}, X., {Xia}, C., \& {Keppens}, R. 2013, \apjl, 771, L29

\bibitem[{{Fang} {et~al.}(2015){Fang}, {Xia}, {Keppens}, \& {Van
  Doorsselaere}}]{fang2015}
{Fang}, X., {Xia}, C., {Keppens}, R., \& {Van Doorsselaere}, T. 2015, \apj,
  807, 142

\bibitem[{{Field}(1965)}]{fiel1965}
{Field}, G.~B. 1965, \apj, 142, 531

\bibitem[{{Froment} {et~al.}(2020){Froment}, {Antolin}, {Henriques},
  {Kohutova}, \& {Rouppe van der Voort}}]{from2020}
{Froment}, C., {Antolin}, P., {Henriques}, V.~M.~J., {Kohutova}, P., \& {Rouppe
  van der Voort}, L.~H.~M. 2020, \aap, 633, A11

\bibitem[{Harten {et~al.}(1983)Harten, Lax, \& Leer}]{hart1983}
Harten, A., Lax, P.~D., \& Leer, B.~v. 1983, SIAM review, 25, 35

\bibitem[{{Hong} {et~al.}(2017){Hong}, {Carlsson}, \& {Ding}}]{hong2017}
{Hong}, J., {Carlsson}, M., \& {Ding}, M.~D. 2017, \apj, 845, 144

\bibitem[{{Johnston} \& {Bradshaw}(2019)}]{john2019b}
{Johnston}, C.~D. \& {Bradshaw}, S.~J. 2019, \apjl, 873, L22

\bibitem[{{Johnston} {et~al.}(2019){Johnston}, {Cargill}, {Antolin}, {Hood},
  {De Moortel}, \& {Bradshaw}}]{john2019a}
{Johnston}, C.~D., {Cargill}, P.~J., {Antolin}, P., {et~al.} 2019, \aap, 625,
  A149

\bibitem[{{Johnston} {et~al.}(2020){Johnston}, {Cargill}, {Hood}, {De Moortel},
  {Bradshaw}, \& {Vaseekar}}]{john2020}
{Johnston}, C.~D., {Cargill}, P.~J., {Hood}, A.~W., {et~al.} 2020, \aap, 635,
  A168

\bibitem[{{Johnston} {et~al.}(2017{\natexlab{a}}){Johnston}, {Hood}, {Cargill},
  \& {De Moortel}}]{john2017a}
{Johnston}, C.~D., {Hood}, A.~W., {Cargill}, P.~J., \& {De Moortel}, I.
  2017{\natexlab{a}}, \aap, 597, A81

\bibitem[{{Johnston} {et~al.}(2017{\natexlab{b}}){Johnston}, {Hood}, {Cargill},
  \& {De Moortel}}]{john2017b}
{Johnston}, C.~D., {Hood}, A.~W., {Cargill}, P.~J., \& {De Moortel}, I.
  2017{\natexlab{b}}, \aap, 605, A8

\bibitem[{{Keppens} {et~al.}(2012){Keppens}, {Meliani}, {van Marle}, {Delmont},
  {Vlasis}, \& {van der Holst}}]{kepp2012}
{Keppens}, R., {Meliani}, Z., {van Marle}, A.~J., {et~al.} 2012, Journal of
  Computational Physics, 231, 718

\bibitem[{{Keppens} \& {Xia}(2014)}]{kepp2014}
{Keppens}, R. \& {Xia}, C. 2014, \apj, 789, 22

\bibitem[{{Klimchuk}(2019)}]{klim2019}
{Klimchuk}, J.~A. 2019, \solphys, 294, 173

\bibitem[{{Linker} {et~al.}(2001){Linker}, {Lionello}, {Miki{\'c}}, \&
  {Amari}}]{link2001}
{Linker}, J.~A., {Lionello}, R., {Miki{\'c}}, Z., \& {Amari}, T. 2001, \jgr,
  106, 25165

\bibitem[{{Lionello} {et~al.}(2009){Lionello}, {Linker}, \&
  {Miki{\'c}}}]{lion2009}
{Lionello}, R., {Linker}, J.~A., \& {Miki{\'c}}, Z. 2009, \apj, 690, 902

\bibitem[{{Meyer} {et~al.}(2012){Meyer}, {Balsara}, \& {Aslam}}]{meye2012}
{Meyer}, C.~D., {Balsara}, D.~S., \& {Aslam}, T.~D. 2012, \mnras, 422, 2102

\bibitem[{{Meyer} {et~al.}(2014){Meyer}, {Balsara}, \& {Aslam}}]{meye2014}
{Meyer}, C.~D., {Balsara}, D.~S., \& {Aslam}, T.~D. 2014, Journal of
  Computational Physics, 257, 594

\bibitem[{{Miki{\'c}} {et~al.}(2013){Miki{\'c}}, {Lionello}, {Mok}, {Linker},
  \& {Winebarger}}]{miki2013}
{Miki{\'c}}, Z., {Lionello}, R., {Mok}, Y., {Linker}, J.~A., \& {Winebarger},
  A.~R. 2013, \apj, 773, 94

\bibitem[{{N{\'o}brega-Siverio} {et~al.}(2020){N{\'o}brega-Siverio},
  {Mart{\'\i}nez-Sykora}, {Moreno-Insertis}, \& {Carlsson}}]{nobr2020}
{N{\'o}brega-Siverio}, D., {Mart{\'\i}nez-Sykora}, J., {Moreno-Insertis}, F.,
  \& {Carlsson}, M. 2020, \aap, 638, A79

\bibitem[{{Parker}(1953)}]{park1953}
{Parker}, E.~N. 1953, \apj, 117, 431

\bibitem[{{Porth} {et~al.}(2014){Porth}, {Xia}, {Hendrix}, {Moschou}, \&
  {Keppens}}]{port2014}
{Porth}, O., {Xia}, C., {Hendrix}, T., {Moschou}, S.~P., \& {Keppens}, R. 2014,
  \apjs, 214, 4

\bibitem[{{Powell} {et~al.}(1999){Powell}, {Roe}, {Linde}, {Gombosi}, \& {De
  Zeeuw}}]{powe1999}
{Powell}, K.~G., {Roe}, P.~L., {Linde}, T.~J., {Gombosi}, T.~I., \& {De Zeeuw},
  D.~L. 1999, Journal of Computational Physics, 154, 284

\bibitem[{{Rempel} \& {Schlichenmaier}(2011)}]{remp2011}
{Rempel}, M. \& {Schlichenmaier}, R. 2011, Living Reviews in Solar Physics, 8,
  3

\bibitem[{{Rempel} {et~al.}(2009){Rempel}, {Sch{\"u}ssler}, \&
  {Kn{\"o}lker}}]{remp2009}
{Rempel}, M., {Sch{\"u}ssler}, M., \& {Kn{\"o}lker}, M. 2009, \apj, 691, 640

\bibitem[{{Ruan} {et~al.}(2020){Ruan}, {Xia}, \& {Keppens}}]{ruan2020}
{Ruan}, W., {Xia}, C., \& {Keppens}, R. 2020, \apj, 896, 97

\bibitem[{{Shu} \& {Osher}(1988)}]{shu1988}
{Shu}, C.-W. \& {Osher}, S. 1988, Journal of Computational Physics, 77, 439

\bibitem[{{Stone} {et~al.}(2020){Stone}, {Tomida}, {White}, \&
  {Felker}}]{ston2020}
{Stone}, J.~M., {Tomida}, K., {White}, C.~J., \& {Felker}, K.~G. 2020, \apjs,
  249, 4

\bibitem[{{Townsend}(2009)}]{town2009}
{Townsend}, R.~H.~D. 2009, \apjs, 181, 391

\bibitem[{{Vaidya} {et~al.}(2017){Vaidya}, {Prasad}, {Mignone}, {Sharma}, \&
  {Rickler}}]{vaid2017}
{Vaidya}, B., {Prasad}, D., {Mignone}, A., {Sharma}, P., \& {Rickler}, L. 2017,
  \mnras, 472, 3147

\bibitem[{van Der~Houwen \& Sommeijer(1980)}]{vand1980}
van Der~Houwen, P.~J. \& Sommeijer, B.~P. 1980, ZAMM-Journal of Applied
  Mathematics and Mechanics/Zeitschrift f{\"u}r Angewandte Mathematik und
  Mechanik, 60, 479

\bibitem[{{van Leer}(1974)}]{vanl1974}
{van Leer}, B. 1974, Journal of Computational Physics, 14, 361

\bibitem[{{van Marle} \& {Keppens}(2011)}]{vanm2011}
{van Marle}, A.~J. \& {Keppens}, R. 2011, Computers and Fluids, 42, 44

\bibitem[{{Vernazza} {et~al.}(1981){Vernazza}, {Avrett}, \&
  {Loeser}}]{vern1981}
{Vernazza}, J.~E., {Avrett}, E.~H., \& {Loeser}, R. 1981, \apjs, 45, 635

\bibitem[{Verwer {et~al.}(1990)Verwer, Hundsdorfer, \& Sommeijer}]{verw1990}
Verwer, J., Hundsdorfer, W., \& Sommeijer, B. 1990, Numerische Mathematik, 57,
  157

\bibitem[{{Withbroe}(1988)}]{with1988}
{Withbroe}, G.~L. 1988, \apj, 325, 442

\bibitem[{{Withbroe} \& {Noyes}(1977)}]{with1977}
{Withbroe}, G.~L. \& {Noyes}, R.~W. 1977, \araa, 15, 363

\bibitem[{{Xia} {et~al.}(2012){Xia}, {Chen}, \& {Keppens}}]{xia2012}
{Xia}, C., {Chen}, P.~F., \& {Keppens}, R. 2012, \apjl, 748, L26

\bibitem[{{Xia} {et~al.}(2011){Xia}, {Chen}, {Keppens}, \& {van
  Marle}}]{xia2011}
{Xia}, C., {Chen}, P.~F., {Keppens}, R., \& {van Marle}, A.~J. 2011, \apj, 737,
  27

\bibitem[{{Xia} \& {Keppens}(2016)}]{xia2016}
{Xia}, C. \& {Keppens}, R. 2016, \apjl, 825, L29

\bibitem[{{Xia} {et~al.}(2017){Xia}, {Keppens}, \& {Fang}}]{xia2017}
{Xia}, C., {Keppens}, R., \& {Fang}, X. 2017, \aap, 603, A42

\bibitem[{{Xia} {et~al.}(2018){Xia}, {Teunissen}, {El Mellah}, {Chan{\'e}}, \&
  {Keppens}}]{xia2018}
{Xia}, C., {Teunissen}, J., {El Mellah}, I., {Chan{\'e}}, E., \& {Keppens}, R.
  2018, \apjs, 234, 30

\bibitem[{{Zhou} {et~al.}(2020){Zhou}, {Chen}, {Hong}, \& {Fang}}]{zhou2020}
{Zhou}, Y.~H., {Chen}, P.~F., {Hong}, J., \& {Fang}, C. 2020, Nature Astronomy

\bibitem[{{Zhou} {et~al.}(2014){Zhou}, {Chen}, {Zhang}, \& {Fang}}]{zhou2014}
{Zhou}, Y.-H., {Chen}, P.-F., {Zhang}, Q.-M., \& {Fang}, C. 2014, Research in
  Astronomy and Astrophysics, 14, 581

\end{thebibliography}

\begin{appendix}
\section{TRACL}
In Section \ref{sec3} we introduced the TRACL method, and we provide further implementation details 
in this appendix. The TRACL methods relies on the magnetic field-line-tracing module in our {\tt MPI-AMRVAC} code, which has been described in Appendix B of \citet{ruan2020}.
Here we document its use for TRACL. While we assume 2D in what follows, this is easily generalized to 3D, the implementation of which is already public domain. 

First of all, we allocated a global array $xy(U, V, ndim)$ that saves the $x$ and $y$ coordinates of all the points on all our field lines, where $U = 2000$ is the maximum number of points allowed for each field line, $V=256$ is the number of field lines used in the TRACL application in this paper, and $ndim = 2$ is for $x$ and $y$, respectively.
Then, we determined the start points of these field lines, which are uniformly distributed along the $y=0$ boundary in this paper, and started to integrate the field lines.
In the following discussion, subscripts $u$ and $v$ mean the value at the $u$-th point on the $v$-th field line, and $m$ and $n$ mean the cell center value in the ($m$, $n$)-th grid cell.
We assumed that the point ($x_{u, v}$, $y_{u, v}$) is located in the upper right corner of the grid cell, whose cell center coordinate is ($x_m$, $y_n$).
Then, the necessary information at this point, for instance, the temperature $T_{u, v}$ and the (in-plane) magnetic field vector $\bm{B}_{u, v}=\left(B_{x,u,v}, B_{y,u,v}\right)$, is interpolated from the neighboring four grid cells ($x_m$, $y_n$), ($x_{m+1}$, $y_n$), ($x_m$, $y_{n+1}$), and ($x_{m+1}$, $y_{n+1}$). This simple bilinear interpolation for
$T_{u, v}$ , for example, gives
\begin{equation}
{T_{u, v}} = \sum\limits_{i = 0}^1 {\sum\limits_{j = 0}^1 {{\alpha _i}{\beta _j}{T_{m + i,n + j}}} },
\label{eqa1}
\end{equation}
where
\begin{eqnarray}
{\alpha _i} = \left| {{x_{u, v}} - {x_{m + 1 - i}}} \right|/\Delta x,
& \,\,\,\,\,\, &
{\beta _j} = \left| {{y_{u, v}} - {y_{n + 1 - j}}} \right|/\Delta y,
\label{eqa3}
\end{eqnarray}
and $\Delta x$ and $\Delta y$ are the grid cell size.
In this way, the position of the next point on the field line ($x_{u+1, v}$, $y_{u+1, v}$) is found from
\begin{equation}
{x_{u + 1, v}} = {x_{u, v}} \pm \Delta l{B_{x, u, v}}/\sqrt {B_{x, u, v}^2 + B_{y, u, v}^2},
\label{eqa4}
\end{equation}
\begin{equation}
{y_{u+1, v}} = {y_{u, v}} \pm \Delta l{B_{y, u, v}}/\sqrt {B_{x, u, v}^2 + B_{y, u, v}^2}.
\label{eqa5}
\end{equation}
$\Delta l$ here is the integration step length, which is chosen as $\Delta l = \min(\Delta x, \Delta y)$ in our simulation.

In the {\tt MPI-AMRVAC} code, the whole domain is divided into blocks, which have a (user-chosen) fixed number of grid cells, as mentioned in Section \ref{sec3}.
Because the AMR grids and blocks evolve dynamically, these blocks may have to be redistributed on the  processors at every time step to achieve load balance.
To manage this, every block is tagged with the current processor id $ipe$ and current block id $iblk$.
The data communication between different processors is basically discussed in \citet{ruan2020}, but a major update to improve the efficiency when interpolating values from field lines back to grid cells is that we allocate two additional global arrays, $ipel(U, V)$ and $iblkl(U, V)$.
These two arrays are used to recode the block id and processor id of every point on every field line.
During the field line integration, according to equation~(\ref{eq17}), the key information of the cutoff temperature $T_{c, v}$ of the $v$-th field line can then be calculated.

To summarize, we then have the cutoff temperature $T_{c}$ for the field line, the $x$ and $y$ position of every point on this line, and information on which processor $ipel$ and in which block $iblkl$ each point is found.
With this information, the cutoff temperature can be interpolated back to the grid cells by
\begin{eqnarray}
{T_{c,m,n}} ={{\sum\limits_v {\sum\limits_u {\frac{{{T_{c,v}}}}{{{{\left( {{x_{u,v}} - {x_{m,n}}} \right)}^2} + {{\left( {{y_{u,v}} - {y_{m,n}}} \right)}^2}}}} } }} \nonumber \\
{{\sum\limits_v {\sum\limits_u {\frac{1}{{{{\left( {{x_{u,v}} - {x_{m,n}}} \right)}^2} + {{\left( {{y_{u,v}} - {y_{m,n}}} \right)}^2}}}} } }}
\label{eqa6}
.\end{eqnarray}
When we know $ipel$ and $iblkl$, the interpolation only needs to be performed over nearby $u$s and $v$s instead of over all points on all field lines.
In this way, the interpolation is much faster than our previous method described in \citet{ruan2020}. With the obtained grid-cell based distribution of the cutoff temperature $T_{c,m,n}$, equations~(\ref{eq18})--(\ref{eq20}) are applied to modify the thermal terms. TRACL strictly follows the 1D TRAC method, but is not so efficient, as the efficiency test in Section \ref{sec4} shows.

\section{TRACB}
Considering that the cutoff temperature $T_c$ does not need to be very accurate, we also introduced the block-based TRAC method, or TRACB method. In the example simulation, we effectively had $1024\times640$ grid cells, that is, $64\times40$ blocks.
Then, we only allocate a global array $T_{c, tab}(M, N)$, where $M=64$ and $N=40$ to save the block-based cutoff temperature.

We know that in a multidimensional MHD simulation, thermal conduction is mostly field-aligned, such that the thermal conduction term $\nabla  \cdot \left( {\kappa  \cdot \nabla T} \right)$ in equation~(\ref{eq6}) is treated as $\nabla  \cdot \left( {{\kappa _\parallel }\left( {\bm{b} \cdot \nabla T} \right)\bm{b}} \right)$, where $\bm{b}$ is the (in-plane) \textbf{unit vector along the magnetic field}. We can hence easily compute the temperature gradient along the field line
and then derive the maximum temperature in a certain block, which fulfills the condition in equation~(\ref{eq17}).
In this way, a single global communication $MPI\_ALLREDUCE$ suffices to obtain the global table $T_{c, tab}$. At the same time, another global table $\bm{b}_{tab}$, which contains the information of all block-center magnetic field vectors, is also communicated to all the processors.

After this global communication, all the processors have the full table $T_{c, tab}$ and $\bm{b}_{tab}$, so that we can integrate the field lines locally in every processor (this computation is hence duplicated on each processor).
Similar to the TRACL method, we selected $V=256$ start points for this integration (so that the computational time can compared fairly), and allocated an $xy_B(W, V, ndim)$ array to save the coordinate information of the points, where $W=400$ is enough for this case.
The interpolation--integration process was the same as discussed above, but we now changed the integration step length to 4$\Delta l$.
Every processor performed the same computations, resulting in the same $xy_B$ so that $T_{c, B}(V)$ tables, collecting cutoff temperature information along field lines.
We could also have divided this integration task over different processors and then accumulated the final result back by an additional global communication, but for the 2D problem here, this was not pursued because the additional communication takes more time than doing it in replicated fashion.

Finally, we interpolated $T_{c, B}$ back into grid cells in the same way as in equation (\ref{eqa6}), but the $xy_B$ array was now sparser than the actual grid point array, therefore we also required the neighboring blocks for this interpolation. To this end, the $iblkl_B(W, V)$ array not only contains the id of the current block, but also the neighboring blocks. After the interpolation, the method works as the TRACL method.
\end{appendix}

\end{document}